\documentclass[12pt,nohyper,letterpaper]{JHEP3}         % For preprint
\usepackage{amssymb,amsfonts}
\usepackage{epsfig}
\def\be{\begin{eqnarray}}
\def\ee{\end{eqnarray}}
\newcommand{\nn}{\nonumber}
\newcommand\para{\paragraph{}}
\newcommand{\ft}[2]{{\textstyle\frac{#1}{#2}}}
\newcommand{\eqn}[1]{(\ref{#1})}

\def\Dslash{\,\,{\raise.15ex\hbox{/}\mkern-12mu D}}
\def\Dbarslash{\,\,{\raise.15ex\hbox{/}\mkern-12mu {\bar D}}}
\def\delslash{\,\,{\raise.15ex\hbox{/}\mkern-9mu \partial}}
\def\delbarslash{\,\,{\raise.15ex\hbox{/}\mkern-9mu {\bar\partial}}}
\def\pslash{\,\,{\raise.15ex\hbox{/}\mkern-9mu p}}
\def\calDslash{\,\,{\raise.15ex\hbox{/}\mkern-12mu {\cal D}}}

\newcommand{\psiy}{\bar{\Psi}_\alpha|\,0\rangle}
\newcommand{\psiyin}{\bar{\Psi}_\alpha\bar{\Psi}_\beta
\bar{\Psi}_{\gamma}|\,0\rangle}

\def\lae{\mathrel{\mathop{\smash{\lower .5 ex \hbox{$\stackrel<\sim$}}}}}
\def\lae{\mathrel{\mathop{\smash{\lower .5 ex \hbox{$\stackrel>\sim$}}}}}

\def\theequation{\thesection.\arabic{equation}}

\def\Dslash{\,\,{\raise.15ex\hbox{/}\mkern-13mu D}}
\def\Dbarslash{\,\,{\raise.15ex\hbox{/}\mkern-12mu {\bar D}}}
\def\delslash{\,\,{\raise.15ex\hbox{/}\mkern-10mu \partial}}
\def\delbarslash{\,\,{\raise.15ex\hbox{/}\mkern-9mu {\bar\partial}}}
\def\pslash{\,\,{\raise.15ex\hbox{/}\mkern-11mu p}}
\def\qslash{\,\,{\raise.15ex\hbox{/}\mkern-9mu q}}
     \def\kslash{\,\,{\raise.15ex\hbox{/}\mkern-11mu k}}
\def\eslash{\,\,{\raise.15ex\hbox{/}\mkern-9mu \epsilon}}
\def\calDslash{\,\,{\rais.15ex\hbox{/}\mkern-12mu {\cal D}}}

\def\bbbone {{\mathchoice {\rm 1\mskip-4mu l} {\rm 1\mskip-4mu
l}{\rm 1\mskip-4.5mu l} {\rm 1\mskip-5mu l}}}
\title{The Geometric Phase and Gravitational Precession of D-Branes}
\author{Chris Pedder, Julian Sonner and David Tong\\
Department of Applied Mathematics and Theoretical Physics, \\
University of Cambridge, UK\\{\tt c.j.pedder, j.sonner,
d.tong@damtp.cam.ac.uk}}

\abstract{We study Berry's phase in the D0-D4-brane system. When a
D0-brane  moves in the background of D4-branes, the first excited
states undergo a holonomy described by a non-Abelian Berry
connection. At weak coupling this is an $SU(2)$ connection over
${\bf R}^5$, known as the Yang monopole. At strong coupling, the
holonomy is recast as the classical gravitational precession of a
spinning particle. The Berry connection is the spin connection of
the near-horizon limit of the D4-branes, which is a continuous
deformation of the Yang and anti-Yang monopole.}

\begin{document}
\pagestyle{plain} \setcounter{page}{1}
\newcounter{bean}
\baselineskip16pt

\section{Introduction}

In a recent paper we explored the Berry phase arising in
supersymmetric quantum mechanics with ${\cal N}=(2,2)$
supersymmetry \cite{goodpaper}. The purpose of this paper is to
extend these results to theories with ${\cal N}=(4,4)$
supersymmetry, specifically the D0-D4-brane system. At weak string
coupling we compute the Berry phase using standard quantum
mechanical techniques; at strong string coupling we use AdS/CFT
methods and show that the quantum holonomy is mapped onto the
classical gravitational precession of a spinning particle.

\para
Berry's phase governs the evolution of a quantum state as the
parameters $\vec{X}$ of the Hamiltonian $H(\vec{X})$ are varied
adiabatically \cite{berry,simon,wz}. We focus on the fate of an
$N$-fold degenerate energy level $|\psi_\alpha\rangle$,
$\alpha=1,\ldots,N$. As the parameters $\vec{X}$ undergo cyclic
evolution in time $T$, the states do not return to themselves but,
instead, undergo a $U(N)$ rotation,
\be |\psi_\alpha(\vec{X}(T))\rangle = \exp\left(-i\int_0^T
E(t)dt\right)\,U_{\alpha\beta}|\psi_\beta(\vec{X}(0))\rangle\ee
The first factor is the dynamical phase. The second factor,
involving the unitary matrix $U$, is the geometric, or Berry,
phase. (For $N>1$, it is more precisely called the Berry
holonomy). $U$ may be expressed as the path-ordered exponential of
the Berry connection $\vec{A}_{\alpha\beta}$,
\be U = P\exp\left(-i\oint \vec{A}\cdot d\vec{X}\right)\ \ \ \
{\rm with}\ \ \
\vec{A}_{\alpha\beta}=i\langle{\psi_\beta}|\frac{\partial}{\partial
\vec{X}}\,|\psi_\alpha\rangle\ee
The canonical example of an Abelian Berry's phase arises for the
ground state of a spin 1/2 particle in a magnetic field $\vec{B}$
\be H = \vec{B}\cdot \vec{\sigma}\label{1st}\ee
where $\vec{\sigma}$ are the Pauli matrices. The Berry connection
for the ground state of this system is famously that of a Dirac
monopole \cite{berry}. The curvature singularity of the monopole
at $\vec{B}=0$ reflects the fact that the two states become
degenerate at this point.

%One follows the ground state $|\Omega(\vec{B})\rangle$, with
%energy $-|\vec{B}|$, as the magnetic field $\vec{B}$ is varied.
%The Berry connection is a $U(1)$ gauge field
%$\vec{A}=-i\langle\Omega|\vec{\nabla}|\Omega\rangle$ defined over
%${\bf R}^3-\{0\}$. A short calculation shows that associated field
%strength is that of the Dirac monopole,
%
%\be \vec{\nabla}\times \vec{A} =
%\frac{\vec{B}}{B^3}\label{dirac}\ee
%
%The Dirac monopole is the unique $SO(3)$ symmetric connection
%first Chern class over ${\bf S}^2 \cong \partial {\bf R}^3$ equal
%to unity. The singularity at $\vec{B}=0$ reflects the fact that
%the two states become degenerate.

\para
There is a natural generalization of \eqn{1st} which yields a
non-Abelian Berry phase. The Hamiltonian is
\be H = \vec{X} \cdot \vec{\Gamma}\label{2nd}\ee
where $\vec{X}$ is a 5-dimensional vector and $\vec{\Gamma}$ are
five $4\times 4$ matrices satisfying the $SO(5)$ Clifford algebra
$\{\Gamma_\mu,\Gamma_\nu\}=2\delta_{\mu\nu}$. The spectrum of
\eqn{2nd} is two-fold degenerate. One may compute the Berry's
phase of the two ground states as $\vec{X}$ is varied
\cite{avron}. This yields an $SU(2)$ connection over ${\bf R}^5$,
known as a {\it Yang monopole} \cite{yang}. An explicit expression
for the Yang monopole is given in equation \eqn{yang}: here we
summarize its main properties.

\para
The Yang monopole can be thought of as an $SU(2)$ generalization
of the Dirac monopole. Both are rotationally invariant
connections, transforming under $SO(3)$ and $SO(5)$ respectively.
Both Dirac and Yang monopoles have their respective Chern class
equal to unity. For the Dirac monopole, this is the first Chern
class, integrated over ${\bf S}^2\cong\partial {\bf R}^3$; for the
Yang monopole the second Chern class integrated over ${\bf
S}^4\cong \partial {\bf R}^5$ is unity. An alternative way of
viewing this is in terms of Hopf maps. The Dirac monopole arises
from the first Hopf map: it is a $U(1)$ fibration over ${\bf S}^2$
to yield ${\bf S}^3$. Similarly, the Yang monopole arises from the
second Hopf map: it is an $SU(2)$ fibration over ${\bf S}^4$
giving ${\bf S}^7$. Finally, like its Abelian counterpart, the
Yang monopole is singular at the origin $\vec{X}=0$. In the
context of Berry's phase, this once again reflects the fact that
all states of the Hamiltonian \eqn{2nd} become degenerate at
$\vec{X}=0$.

\para
The Yang monopole first appeared as a Berry phase in the context
of time-reversal invariant fermi systems \cite{avron}. It has
since found applications in the $SO(5)$ theory of
antiferromagnetism and high $T_c$ superconductivity
\cite{bigzhang}, spin currents in superconductors
\cite{middlezhang}, and spin pairing in ultra-cold atomic systems
\cite{smallzhang}. The quaternionic structure of the connection
were elaborated upon in \cite{levay} (see also \cite{levay2} for
further developments). Finally, the emergence of this Berry
connection in D-brane systems was discussed previously in the
context of the $Sp(N)$ matrix model for type I string theory
\cite{usp}, work which has some overlap with the discussion in
this paper.

\para
In Section 2 of this paper we study the Berry connection in
D-brane systems, specifically in the quantum mechanics of the
D0-D4 system. This system has a unique ground state. More
interesting are the first excited states carrying quantum numbers
under the R-symmetry of the theory. At weak string coupling we
compute the non-Abelian Berry phase for these states as the
D0-brane moves in the background of the D4-branes. We show that
the Berry connection consists of copies of the Yang and anti-Yang
monopoles. To our knowledge, this is the first study of Berry's
phase in systems with 8 supercharges, and the resulting Yang
monopole connection differs from those that appeared in previous
studies of Berry's phase in ${\cal N}=(2,2)$ supersymmetric
quantum mechanics as chiral multiplet \cite{ttstar1,ttstar2} and
vector multiplet \cite{goodpaper} parameters are varied.

\para
In Section 3 we study the strong coupling limit of the quantum
mechanics. Here the relevant description is in terms of a probe
D0-brane moving in the near-horizon geometry of the D4-branes, and
the R-symmetry quantum numbers have the interpretation of the spin
of the particle in the gravitational background. As the particle
moves, this spin undergoes classical gravitational precession,
providing the manifestation of Berry's phase at strong coupling.
We show that the gravitational spin connection is a continuous
deformation of the weak-coupling Yang-anti-Yang monopole Berry
connection.

\section{Quantum Mechanics}

The quantum mechanics describing the D0-D4-system was discussed
previously [11-17].
%\cite{dkps,diac,evaabs,abs,berkver,ss,ground}.
The theory  admits ${\cal N}=(4,4)$ supersymmetry\footnote{The
notation reflects the fact that the superalgebra is the
dimensional reduction of the vector-like ${\cal N}=(4,4)$ algebra
in two dimensions. It has 8 real supercharges and is sometimes
called ${\cal N}=8A$, or simply ${\cal N}=8$, supersymmetry.} and
descends from the reduction of ${\cal N}=1$ supersymmetric
theories in $d=5+1$ dimensions. The R-symmetry group of the
quantum mechanics is
\be R = Spin(5) \times SU(2)_R \cong Sp(2) \times Sp(1)\ee
The $Spin(5)$ symmetry is the remnant of the $SO(9,1)$ Lorentz
symmetry transverse to the $D4$-branes; $SU(2)_R$ is part of the
spatial $SO(4)\cong SU(2)_L\times SU(2)_R$ symmetry of the
D4-brane worldvolume. The four complex supercharges $Q_\alpha$,
$\alpha=1,2,3,4$ transform in the ${\bf 4}$ of $Spin(5)$ while
$(Q,JQ^\star)$ form a doublet of $SU(2)_R$, where $J^2=-1$. (This
latter transformation follows from the fact that the supercharges
form a symplectically real chiral spinor in six dimensions).

\para
The massless representations of the superalgebra include the
familiar hypermultiplet and vector multiplet. The hypermultiplet
contains two complex scalars $\phi$ and $\tilde{\phi}$. These are
singlets under $Spin(5)$ while $(\phi,\tilde{\phi}^\dagger)$
transforms as a doublet under $SU(2)_R$. There are also four
complex fermions $\Psi_\alpha$, $\alpha=1,2,3,4$ transforming as
$({\bf 4},{\bf 1})$. The vector multiplet contains five real
scalars $\vec{X}$ transforming as $({\bf 5},{\bf 1})$ under $R$,
together with four complex fermions $\Lambda_\alpha$ which
transform as $({\bf 4},{\bf 2})$. Finally, there is a
non-dynamical gauge field $v$ whose role is to impose constraints
so that the Hilbert space consists of charge-neutral states.

\subsection{The Lagrangian}

We will ignore the trivial motion in the direction parallel to the
D4-branes, and their fermionic partners which result in a 16-fold
degeneracy of all states. The low-energy dynamics of the D0-brane
is described by $U(1)$ gauged quantum mechanics coupled to $N$
charged hypermultiplets arising from quantizing the $D0-D4$
strings \cite{berkdoug}. The Lagrangian takes the form
\be L= L_{vector} + L_{hyper} + L_{Yuk} \label{lag}\ee
The first term describes the free motion of the D0-brane in the
directions transverse to the D4-brane.
\be L_{vector} = \frac{1}{2g^2}\, (\dot{\vec X}{}^2 + 2i
\bar{\Lambda}\dot{\Lambda})\label{lagvec}\ee
Here we are using conventions typical of higher dimensional gauge
theories, in which the scaling dimensions are given by $[g^2]=3$
and $[\vec{X}]=1$. In terms of string theory parameters,
$g^2=g_s/(2\pi)^2\alpha^{\prime\,3/2}$ while the distance between
the D0-brane and D4-branes is $2\pi\alpha^\prime\vec{X}$. The mass
of the D0-brane is $M_{D0}=1/g_s\sqrt{\alpha^\prime}$. The
decoupling limit for the D0-brane quantum mechanics requires us to
keep both $g^2$ and $\vec{X}$ fixed, while sending
$g_s\rightarrow 0$ and $\alpha^\prime \rightarrow 0$.

\para
The hypermultiplet scalars $\phi_i$ and $\tilde{\phi}_i$,
$i=1,\ldots, N$ have charge $+1$ and $-1$ respectively. The
fermions $\Psi_{\alpha i}$ each have charge $+1$. Their
interactions are given by\footnote{The canonical conventions for
two-dimensional supersymmetric gauge theories were presented in
\cite{phases}. Our Lagrangian follows upon dimensional reduction
to quantum mechanics, with the substitution $\Psi_\alpha=
(\psi_+,\psi_-,\bar{\tilde{\psi}}_+,\bar{\tilde{\psi}}_-)$ and
$\Lambda_\alpha =
\left(-i\bar\lambda_-\,,-i\bar\lambda_+\,,\eta_-\,,\eta_+
\right)$.}
\be L_{hyper} &=& \sum_{i=1}^N |{\cal D}_t\phi_i|^2 + |{\cal D}_t
\tilde{\phi}_i|^2+i\bar{\Psi}_i{\cal D}_t\Psi_i -
\vec{X}^2(|\phi_i|^2 + |\tilde{\phi}_i|^2) \nn\\ && -
\frac{g^2}{2} (\sum_i |\phi_i|^2 - |\tilde{\phi}_i|^2)^2 - 2g^2
|\sum_i \tilde{\phi}_i\phi_i|^2 \ee
The $\vec{X}^2|\phi|^2$ terms reflect the mass of the stretched
string when the D0-brane lies a distance $\vec{X}$ from the
D4-brane. Finally, the Yukawa interactions between fermions are
given by
\be L_{Yuk} = -\bar{\Psi}\,(\vec{X}\cdot\vec{\Gamma})\,\Psi +
\sqrt{2}\bar{\Psi}_\alpha(\phi\Lambda_\alpha +
\tilde{\phi}^\dagger J_\alpha^{\
\beta}\Lambda^\star_\beta) + \rm{h.c.}\label{lagyuk}\ee
with $J$ a $4\times 4$ symplectic matrix such that $J^2 = -1$,
while $\vec{\Gamma}$ furnish a representation of the $SO(5)$
Clifford algebra: $\{\Gamma_a,\Gamma_b\}= 2 \delta_{ab}$. An
explicit representation of these matrices is given in Appendix 1.

\subsection{The Born-Oppenheimer Approximation}
\label{sec:openstring}

The full quantum mechanics \eqn{lag} is rather complicated. For
example, an analysis of the ground-state wavefunction may be found
in \cite{ss,ground}. We work in the Born-Oppenheimer approximation
in which we freeze the D0-brane at a fixed position $\vec{X}$ and
quantize the hypermultiplets associated to the D0-D4 strings. This
is valid in the regime $g^2 \ll X^3/N$, where $X=|\vec{X}|$. Here
we can neglect the motion of the slow, heavy D0-brane and focus
attention on the fast hypermultiplets. We will then be interested
in the Berry's connection for the states as we vary $\vec{X}$ and
the D0-brane moves adiabatically in the background of the
D4-brane.

\para
We start our discussion by considering the case of a single
hypermultiplet, $N=1$, in the strict limit $g^2=0$. We will return
to the case of finite $g^2$ in Section \ref{ccc}. The quantum
mechanics of a single hypermultiplet in the Born-Oppenheimer
approximation is simply given by the free Lagrangian
\be L_{\rm qm} = |\dot\phi|^2 +  |\dot{\widetilde{\phi}}|^2  + i
\overline{\Psi}\dot\Psi - X^2
 \left(|\phi|^2 + |\tilde\phi|^2  \right) - \bar{\Psi}
 \left( \vec{X}\cdot \vec{\Gamma} \right)\Psi
\ee\label{easy}
Despite its simplicity, this Lagrangian already contains
interesting physics. This arises, most notably, from the fermion
mass term which, the reader will note, is similar in form to the
Hamiltonian \eqn{2nd} discussed in the introduction.

\para
We pass to the Hamiltonian formalism by introducing the canonical
momenta $\pi=\dot{\phi}^\dagger$,
$\tilde{\pi}=\dot{\tilde{\phi}}{}^\dagger$ and the fermionic
conjugate momenta $\partial L/\partial \dot{\Psi}=-i\bar{\Psi}$,
giving us a Hamiltonian consisting of $4$ bosonic and $4$
fermionic harmonic oscillators,
\be H=|\pi|^2 + |\tilde{\pi}|^2 + X^2
\left(|\phi|^2+|\tilde\phi|^2  \right) + \bar{\Psi} \left(
\vec{X}\cdot \Gamma \right)\Psi \label{simpleham}\ee
We construct the Hilbert space from the fermionic operators in the
usual fashion. The canonical anti-commutation relations read
$\{\bar{\Psi}_\alpha,\Psi_\beta\}=\delta_{\alpha\beta}$. We define
the reference state $| 0 \rangle$ to be annihilated by
$\Psi_\alpha |\,0\rangle = 0$. Then we form a basis of the
fermionic Hilbert space ${\cal H}^F$ by acting on $|\,0\rangle$
with the creation operators $\bar{\Psi}$ to form a tower of
$2^4=16$ states as shown in the table. The multiplicity of states,
together with their eigenvalues under the fermionic Hamiltonian
\be H_F=\bar{\Psi}(\vec{X}\cdot\vec{\Gamma})\Psi\label{westham}\ee
are also shown. The subscripts on the energy eigenvalues denote
the degeneracy of the state.
\begin{center}
\begin{tabular}{|c|c|c|} \hline
State & Multiplicity & $H_F$ Eigenvalue
\\ \hline
$|\,0\rangle$ & 1 &  0 \\ \hline $\bar{\Psi}_\alpha |\,0\rangle$ &
4 & $(-X)_2$ , $(+X)_2$ \\ \hline
$\bar{\Psi}_\alpha \bar{\Psi}_\beta|\,0\rangle$ & 6  & $-2X$, $0_4$, $+2X$\\
\hline $\bar{\Psi}_\alpha\bar{\Psi}_\beta\bar{\Psi}_\gamma
|\,0\rangle$ & 4 & $(-X)_2$, $(+X)_2$\\ \hline
$\bar{\Psi}_\alpha\bar{\Psi}_\beta\bar{\Psi}_\gamma
\bar{\Psi}_\delta|\,0\rangle$ & 1 & 0 \\ \hline \end{tabular}
\\
[0.3cm]Table 1: The Fermionic Hilbert Space.
\end{center}

\subsubsection*{The Ground State}

Our quantum mechanical system has a unique ground state
$|\Omega\rangle$. (Indeed, the full Lagrangian \eqn{lag} is
conjectured to have a unique ground state. This has been proven
for the case $N=1$ of a single D4-brane \cite{ss,ground}). For our
truncated, free model the fermionic part of the ground state
$|\Omega\rangle$ is a linear combination of the six states
$\bar{\Psi}_\alpha \bar{\Psi}_\beta|\,0\rangle$ such that $H_F$
has eigenvalue $-2X$. The complex bosonic fields $\phi$ and
$\tilde{\phi}$ are placed in their Gaussian vacuum state, each
contributing $+X$ to the energy. Thus the total energy is
$H|\Omega\rangle = (-2X+X+X)|\Omega\rangle =0$, as expected for a
supersymmetric system.

\para
The ground state $|\Omega\rangle$ depends on $\vec{X}$ or, more
precisely, on the direction $\vec{X}/X$, through its presence in
the fermionic Hamiltonian $H_F$. One can compute the Abelian
Berry's phase for the ground state as $\vec{X}$ is varied. This
defines a $U(1)$ connection over ${\bf R}^5-\{0\}$
\be \vec{A} = i\langle\Omega|\frac{\partial}{\partial
\vec{X}}|\Omega\rangle \ee
A straightforward calculation shows that this connection is
trivial. There is no holonomy in the phase of the ground state
wavefunction as the D0-brane moves around the
D4-branes\footnote{There is, however, a closely related situation
where the D0-D4-brane system does admit a Berry's phase in the
ground state. Consider the case of $k$ D0-branes. In the limit
$k\rightarrow \infty$, there is a BPS configuration of the
D0-brane theory with $[X^1,X^2]\sim 2\pi i$, describing a D2-brane
extended transverse to the D4-brane. The Berry phase of this
ground state is the Dirac monopole \cite{berkdoug}, reflecting the
fact that the D2-brane is magnetically charged under the D4-brane
RR gauge field. This calculation is closely related to Berry's
phase in  ${\cal N}=(2,2)$ quantum mechanics \cite{goodpaper}.
Related Aharonov-Bohm effects for branes were also recently
discussed in \cite{sean}. }.

\subsection{Excited States and Non-Abelian Berry's Phase}
\label{sec:su2berry}

We now turn to the excited states. At the first level there are
four degenerate states, arising from the eigenvectors of $H_F$
with eigenvalue $-X$. Once dressed with the bosonic ground states,
each of these has energy $+X$, as befits a stretched string
between the D0-brane and D4-brane.

\para
The four excited states split naturally into two pairs. One pair
lives in the $\bar{\Psi}_\alpha|\,0\rangle$ sector, while the
other lives in the
$\bar{\Psi}_\alpha\bar{\Psi}_\beta\bar{\Psi}_{\gamma}|\,0\rangle$
sector. In our free theory \eqn{easy} there is no mixing between
these states. Let us start by focussing attention on the first of
these pairs. The bosons $\phi$ and $\tilde{\phi}$ are placed in
their Gaussian ground state; this depends on the magnitude, but
not the direction, of $\vec{X}$ which ensures that the bosons will
not affect Berry's phase. We disregard the bosons for now, but
will return to them shortly. The fermionic states
$\bar{\Psi}_\alpha|\,0\rangle$ form a basis of the
four-dimensional Hilbert space ${\cal H}_4$. The fermionic
Hamiltonian \eqn{westham} on this space acts as
\be H_F=\vec{X}\cdot\vec{\Gamma}\label{eastham}\ee
We define the projection operators
\be P_{\pm} = \frac{1}{2}\left(\bbbone \pm
\frac{\vec{X}}{X}\cdot\vec{\Gamma} \right) \ee
The pair of states of interest lie in the two-dimensional
eigenspace $P_-{\cal H}_4$. We are interested in the holonomy of
these states as the D0-brane moves adiabatically in the background
of the D4-brane. This is given in terms of an (a priori) $U(2)$
connection over ${\bf R}^5$. For each $\vec{X}$, we pick an
arbitrary basis $\{|\,1\rangle, |\,2\rangle\}$ of $P_-{\cal H}_4$.
Then the non-Abelian Berry connection is defined by
\be (A_\mu)_{ab} = i\langle a |\frac{\partial}{\partial X^\mu}
|\,b\rangle \ee
and, under the adiabatic evolution of $\vec{X}$, the
transformation of the state includes both the dynamical phase and
the geometrical Berry phase, the latter given by
\be |\,a\rangle \rightarrow P\,\exp\left(-i\oint (A_{\mu})_{ab}
dX^\mu\right)\,|\,b\rangle\ee
To give an explicit expression for the Berry connection, we must
first choose a gauge which, in this context, means picking a basis
of states spanning $P_-H_F$. We choose the basis of
(un-normalized) states,
\be |\,1\rangle = P_-\bar{\Psi}_1|\,0\rangle \ \ {\rm and}\ \ \
|\,2\rangle = P_- \bar{\Psi}_3|\,0\rangle\ee
which are valid everywhere except along the $\vec{X}=(0,0,0,0,-X)$
axis where the ground states are orthogonal to both
$\bar{\Psi}_1|\,0\rangle$ and $\bar{\Psi}_3|\,0\rangle$. As a
result, the Berry connection we derive will have a Dirac string
singularity along this axis. The explicit computation of the Berry
connection is relegated to Appendix B. The result of the
computation is,
\be (A_\mu)_{ab} =
\frac{-X^\nu}{2X(X+X_5)}\,\eta^m_{\mu\nu}\,\sigma^{m}_{ab} \qquad
\mu = 1,2,3,4 \ \ \ \ ,\ \ \ \ A_5 =0 \label{yang}\ee
where $\eta_{\mu\nu}^m$ are the 't Hooft matrices defined in
Appendix A and $\sigma_{ab}^m$ are the Pauli matrices.

\para
This is the connection for the Yang monopole. The singularity
along the positive $X^5$ axis is merely a gauge artifact. In
contrast, the singularity at the origin $\vec{X}=0$ is real and
reflects the fact that the states with $H_F$ eigenvalue $\pm X$
become degenerate at this point. The defining property of this
connection is that the field strength $F_{\mu\nu}= \partial_\mu
A_\nu -
\partial_\mu A_\nu - i[A_\mu,A_\nu]$ has second Chern class
\be c_2=\frac{1}{8\pi^2}\int_{{\rm S}^4}{\rm tr}\left( F\wedge
F\right) = -1 \ee
when evaluated over any ${\bf S}^4$ centered at the origin.

\subsubsection*{Yin and Yang}

It is a simple matter to repeat this analysis for the equivalent
states in the
$\bar{\Psi}_\alpha\bar{\Psi}_\beta\bar{\Psi}_{\gamma}|\,0\rangle$
sector. In the basis
$\epsilon_{\alpha\beta\gamma\delta}\bar{\Psi}_\beta
\bar{\Psi}_\gamma\bar{\Psi}_\delta |\,0\rangle$, the fermionic
Hamiltonian \eqn{westham} takes the form
\be H_F=-\vec{X}\cdot\vec{\Gamma}^* \equiv - \vec{X}\cdot
J^{-1}\vec{\Gamma}J \ee
where $J$ is the charge conjugation matrix
$J=-\Gamma^3\Gamma^4=\tiny{\left(\begin{array}{cc} 0 & -1 \\ 1 & 0
\end{array}\right)}$. It will prove convenient to work in the
alternative basis $(J^{-1})^{\
\beta}_{\alpha}\epsilon_{\beta\gamma\delta\lambda}\bar{\Psi}_\gamma
\bar{\Psi}_\delta\bar{\Psi}_\lambda |\,0\rangle$, in terms of
which the fermionic Hamiltonian \eqn{westham} has matrix elements
\be H_F = -\vec{X}\cdot \vec{\Gamma}\ee
This has the same form as \eqn{eastham}, but the extra minus sign
ensures that states of interest are now those in the eigenspace
$P_+{\cal H}_4$. A similar computation to that above, reviewed in
Appendix B, shows that the $SU(2)$ Berry connection $\tilde{A}$ in
this sector is the anti-Yang monopole, or {\it Yin monopole}, with
second Chern class $c_2=+1$:
\be (\tilde{A}_\mu)_{ab} = \frac{-X^\nu}{2X(X+X_5)}
\,\bar\eta^m_{\mu\nu}\,\sigma^{m}_{ab}\qquad \mu = 1,2,3,4 \ \ \ \
,\ \ \ \ \tilde{A}_5 =0 \label{yin}\ee
where the anti-self dual 't Hooft matrices $\bar{\eta}$ are given
in Appendix A.

\para
Including both $\bar{\Psi}_\alpha |0\rangle$ and
$\bar{\Psi}_\alpha\bar{\Psi}_\beta\bar{\Psi}_{\gamma}|\,0\rangle$
sectors, the Berry's phase for the four first excited states is a
reducible Yin-Yang monopole. This is a $4\times 4$ block-diagonal
matrix,
\be \omega_\mu = \left(\begin{array}{cc}\tilde A_\mu & 0 \\
0 & A_\mu \end{array}\right)\ \ \ \mu=1,2,3,4\ \ \ \ \ , \
\ \ \ \ \  \omega_5=0\label{yy0}\ee
In the following section we will compute the Berry connection in
the strong coupling regime using AdS/CFT. To facilitate
comparison, it will be useful to perform a gauge transformation to
a more symmetric gauge. We rotate the states using the singular
gauge transformation \cite{smallzhang}
\be U = V \  \exp\left(\frac{i\theta_1X^\nu\Gamma_{\nu
5}}{\sqrt{X^2-(X^5)^2}}\right) \ee
where $X^5=X\cos\theta_1$ and
\be V =  \scriptsize{\left(\begin{array}{cccc}
1 & 0 & 0 & 0\\
0 & 0 & 1 & 0\\
0 & 1 & 0 & 0\\
0 & 0 & 0 & 1  \end{array} \right)}  \ee
The reducible Berry connection
\eqn{yy0} transforms as $\omega_\mu\rightarrow U\omega_\mu
U^\dagger  - i(\partial_\mu U) U^\dagger$, after which it takes
the more symmetric form
\be \omega_\mu=\frac{X^\nu}{X^2}\Gamma_{\mu\nu}\ \ \ \
\mu,\nu=1,\ldots,5\label{yy}\ee
where $\Gamma_{\nu\mu}=\ft{1}{4i}[\Gamma_\nu,\Gamma_\mu]$ are the
spinor generators of $Spin(5)$.

\subsection{Caveats, Constraints and Complications}
\label{ccc}

The above calculation is valid in the strict $g^2 \rightarrow 0$
limit, ensuring that the vector multiplet fields are frozen.
Typically, when performing a Born-Oppenheimer approximation, there
is no subtlety in subsequently turning on kinetic terms for the
heavy degrees of freedom $\vec{X}$. However, in our supersymmetric
theory turning on finite $g^2$ necessarily involves also turning
on terms other than kinetic terms for $\vec{X}$. In particular,
there are effects due to the vector multiplet fermions $\Lambda$
and the gauge field $v$. In this section we study how these, and
other complications, affect the Berry's phase calculation.

\subsubsection*{Gauss' Law}
\label{gauss}

In the calculation of Section \ref{sec:su2berry} we have neglected
the effect of the gauge field. While this is valid in the strict
$g^2\rightarrow 0$ limit, at finite $g^2$ we must treat it
correctly. In quantum mechanics, the gauge field is
non-propagating: rather it imposes the constraint of Gauss' law on
states, restricting the Hilbert space to the charge-neutral
sector.

\para
The unique ground state of our system lies in the sector
$\bar{\Psi}_\alpha\bar{\Psi}_\beta|\,0\rangle$. We require that
this ground state survives the projection of Gauss' law. Since
each $\bar{\Psi}_\alpha$ has charge $-1$, we must endow the
reference state $|\,0\rangle$ with charge $+2$.

\para
Let us now look at the first excited states in the sector
$\bar{\Psi}_\alpha|\,0\rangle$. These states have charge $+1$ and
do not survive the Gauss purge. To rectify the situation, we must
turn to the bosons which, until now, have been languishing in
their ground state. It is time to shake them out of their lethargy
and put them to work. We have two complex bosons, $\phi$ and
$\tilde{\phi}$, carrying charge $+1$ and $-1$ respectively. At
lowest order, we therefore have two possibilities to create a
charge neutral state: we may either excite a single quantum of
$\phi^\dagger$, or a single quantum of $\tilde{\phi}$. This is
done by the usual creation operators of the complex harmonic
oscillator,
\be \bar{a}^\dagger &=& \frac{1}{\sqrt{2X}}(X\phi^\dagger - i\pi)
\nn\\ \bar{\tilde{a}} &=& \frac{1}{\sqrt{2X}}(X\tilde{\phi} -
i\tilde{\pi}^\dagger) \ee
The effect of imposing charge neutrality is a doubling of the
lowest lying states. For each of the two lowest lying states in
the $\bar{\Psi}_\alpha|\,0\rangle$ sector, we may act with either
$\bar{a}^\dagger$, or $\tilde{a}^\dagger$, resulting in 4 charge
neutral states,
\be
 \left(\begin{array}{c}
\bar a^\dagger \\
\bar{\widetilde{a}}
\end{array}\right)\Psi_\alpha^\dagger|\,0\rangle\qquad
\label{su2doub}\ee
each of which has energy $2X$. This can also be understood from
the string-theory picture, where a single string ending on a
D0-brane suffers a tadpole. In order to cancel this, another
string must leave the D0-brane. Hence, these states correspond to
a string and anti-string stretched from the D4-brane to the
D0-brane. This fact will be important later. Note also that the
bosonic creation operators sit in an $SU(2)_R$ doublet. The
resulting states therefore carry spin in the directions parallel
to the D4-branes.

\para
The story for the lowest excited states in the $\psiyin$ sector is
similar. These states have charge $-1$ and must be dressed with a
either single quantum of $\phi$ or $\tilde{\phi}^\dagger$. This is
achieved by acting with the creation operators,
\be \bar{a}&=&\frac{1}{\sqrt{2X}}(X\phi-i\pi^\dagger) \nn\\
\bar{\widetilde{a}}^\dagger &=&
\frac{1}{\sqrt{2X}}(X\tilde{\phi}^\dagger-i\tilde{\pi})\ee
This again results in four states, each with energy $2X$, sitting
in two doublets of $SU(2)_R$,
\be \left(\begin{array}{c}
\bar{a} \\
\bar{\widetilde{a}}^\dagger
\end{array}\right)\epsilon_{\alpha\beta\gamma\delta}\Psi_\beta^\dagger
\Psi_\gamma^\dagger\Psi_\delta^\dagger|\,0\rangle \ee
The bosonic contributions to the wavefunctions depend only on $X$;
they have no dependence on the angular part of $\vec{X}$. This
ensures that they do not contribute to Berry's phase. Similarly,
the states of energy $2X$ which lie in the
$\bar{\Psi}_\alpha\bar{\Psi}_\beta|\,0\rangle$ sector (for
example,
$\bar{a}\bar{\tilde{a}}\bar{\Psi}_\alpha\bar{\Psi}_\beta|\,0\rangle$)
have no Berry's phase and do not mix with the other sectors.

\para
The upshot of imposing Gauss' law is that the there are now eight
excited states of interest, each with energy $2X$. The Berry's
phase consists of an $SU(2)_R$ doublet of the Yin-Yang monopole
\eqn{yy0}.

\subsubsection*{Adding Flavour}

The above computations are for a single D4-brane. The
generalization to $N>1$ D4-branes is straightforward. In the
$g^2=0$ limit, there are no interactions between different
sectors. We have $N$ copies of the pair of excited states in the
$\bar{\Psi}_{i\alpha}|\,0\rangle$ sector, with $i=1,\ldots, N$,
and, correspondingly, $N$ copies of the Yang monopole connection.
There is an $SU(N)$ flavour symmetry rotating these states.

\para
We choose to focus on the $SU(N)$ singlet sector. The simplest
singlet state is constructed as follows. We first build a charge
$+1$ state transforming in the ${\bf N}$ of $SU(N)$. To achieve
this, we place all but one of the flavours in their unique ground
state, living in the sector
$\bar{\Psi}_{j\alpha}\bar{\Psi}_{j\beta}|\,0\rangle$, $j\neq i$.
The remaining flavour sits in its lowest excited state in the
sector $\bar{\Psi}_{i\gamma}|\,0\rangle$. This gives rise to an
$N$-tuplet of pairs of excited states, each of energy $+X$.
Schematically they sit in the sector
\be \bar{\Psi}_{i\gamma} \prod_{j\neq
i}\bar{\Psi}_{j\alpha}\bar{\Psi}_{j\beta}\ |\,0\rangle\ ,\ \ \ \ \
i=1,\ldots, N \ee
These states are neither charge- nor flavour-neutral. We now tensor
this fermionic state with the bosonic excitation $(\bar
a_i^\dagger, \bar{\widetilde{a}}_i)$, which has charge $-1$ and
transforms in the $\bar{\bf N}$ of $SU(N)$. This results, as in
Section \ref{gauss} in four degenerate states, each of energy
$+2X$. They are $SU(N)$ singlets, and form two pairs of $SU(2)_R$
doublets, each transforming under the Yang monopole connection. A
similar construction works for the Yin monopole in the $\psiyin$
sector.

\subsubsection*{Fine Structure and Symmetries}

In addition to the scalar fields $\vec{X}$, we have their
fermionic partners $\Lambda_\alpha$. These could be happily
ignored in the  $g^2\rightarrow 0$, but when working at finite
$g^2$ we must take them into account. They have two effects.
Firstly they introduce new fermionic states in the quantum theory.
Secondly, the Yukawa terms cause a fine structure splitting of
previously degenerate states.

\para
Upon quantizing the Lagrangian \eqn{lagvec}, we have the
anti-commutation relations
\be \{\Lambda_\alpha,\bar{\Lambda}_\beta\} =
g^2\,\delta_{\alpha\beta}\ee
We choose our reference state to satisfy $\Psi_\alpha|\,0\rangle =
\Lambda_\alpha|\,0\rangle = 0$. Dressing our states with
$\bar{\Lambda}$ operators then provides a 16-fold degeneracy for
each state. This degeneracy is subsequently lifted by the Yukawa
coupling,
\be L_{Yuk} = \sqrt{2}\bar{\Psi}_\alpha(\phi\Lambda_\alpha +
\tilde{\phi}^\dagger J_\alpha^{\ \beta}\Lambda_\beta^\star) + {\rm
h.c.}\label{luyk}\ee
Here $J$ is the $4\times 4$ symplectic matrix with block form
$J=\tiny{\left(\begin{array}{cc} 0 & -1 \\ 1 & 0
\end{array}\right)}$. It is worth pausing to note how this term is
invariant under the $Spin(5)\times SU(2)_R$ R-symmetry group. Both
$\Psi$ and $\Lambda$ transform under the ${\bf 4}$ of $Spin(5)$.
However, $\Psi$ is an $SU(2)_R$ singlet, while
$(\phi,\tilde{\phi}^\dagger)$ forms an $SU(2)_R$ doublet. We see
that $SU(2)_R$ invariance is maintained if $(J\Lambda^\star,
\Lambda)$ transform as a doublet.

\para
The Yukawa couplings \eqn{luyk} only preserves fermion number
modulo 2. At leading order in $g^2$ our excited states in the
$\psiy$ sector and
$\bar{\Psi}_\alpha\bar{\Psi}_\beta\bar{\Psi}_\gamma|\,0\rangle$
sector mix with states of the form
$\bar{a}\bar{\tilde{a}}\bar{\Lambda}_\alpha\bar{\Psi}_\beta\bar{\Psi}_\gamma
|\,0\rangle$, which also have energy $2X$. We have not computed
the fine structure splitting arising from the mixing of our
$8\times 16$ states with the associated states in the
$\bar{\Lambda}_\alpha\bar{\Psi}_\beta\bar{\Psi}_\gamma |0\rangle$
sector. Here we merely note that, in the full theory, the states
must form representations of the $Spin(5)_R\times SU(2)_R$
symmetry. The quantum numbers of our states of interest are
imprinted by $\Lambda$ with which they mix: the states transform
in $({\bf 4},{\bf 2})$ representation.

\section{Gravitational Precession}

In this section we study the D0-D4 system as $g_s$ is increased.
We will study the Berry's phase for the lowest $U(N)$ singlet
state transforming in the $({\bf 4},{\bf 2})$ of the
$Spin(5)\times SU(2)_R$ R-symmetry. We start by studying the
strong coupling limit of the quantum mechanics; we then make some
comments about the decoupling limit of the D4-brane.

\subsection{Strong Coupling in the Quantum Mechanics}

While the Born-Oppenheimer calculation described in Section 2 is
valid at $g^2N \ll \langle X\rangle^3$, we may also study the
quantum mechanics in the opposite limit $g^2N\gg \langle X\rangle
^3$. This is the infra-red regime of  the quantum mechanics, where
the Coulomb branch and the Higgs branch decouple \cite{evaabs}. We
will be interested in the quantum mechanics of the Higgs branch,
reflecting our interest in the $D0-D4$ strings in Section 2.

\para
Although we are interested in the Higgs branch quantum mechanics,
the most useful variables to describe the flavor singlet sector of
the Higgs branch are actually those of the Coulomb branch. At the
classical level, this relationship follows simply from setting
$g^2=\infty$ in the Lagrangian, resulting in the algebraic
equation for $\vec{X}$ in terms of fermi bi-linears
\cite{berkver}\footnote{This map is a supersymmetric version of
that employed to solve the Gross-Neveu model \cite{gn}. It also
provides a useful change of coordinates in $d=0+0$ \cite{dhkmv}
and $d=1+1$ \cite{aber} supersymmetric theories.}.
\be
\vec{X}=\frac{\sum_i\bar{\Psi_i}\vec{\Gamma}\Psi_i}{\sum_j\,|\phi_j|^2+
|\tilde{\phi}_j|^2}\label{map}\ee
To derive an expression for the Higgs branch dynamics at strong
coupling, one does something a little unintuitive: we integrate
out the hypermultiplets, in their {\it ground state}, at weak
coupling $g^2N \ll X^3$. At first glance, neither of these things
seems to make much sense because we are interested in the {\it
excited state} of the object we're  integrating out at {\it
strong} coupling! Nonetheless, as we now explain, this turns out
to be the correct strategy.

\para
Integrating out the hypermultiplets gives rise to an effective
action for the Coulomb branch quantum mechanics. The ${\cal
N}=(4,4)$ supersymmetry prevents the generation of a potential on
the Coulomb branch. The first non-trivial corrections to the
bosonic effective action may be simply computed using a background
field method, in which $X\rightarrow X +\delta X$. They arise from
the following diagrams: the external dashed lines are the
${\vec{X}}$ of frequency $\pm k$; the internal solid lines are
either $\phi_i$ or $\tilde{\phi}_i$ for $i=1,\ldots,N$. Each
contributes
\be \raisebox{-3.2ex}{\epsfxsize=1.4in\epsfbox{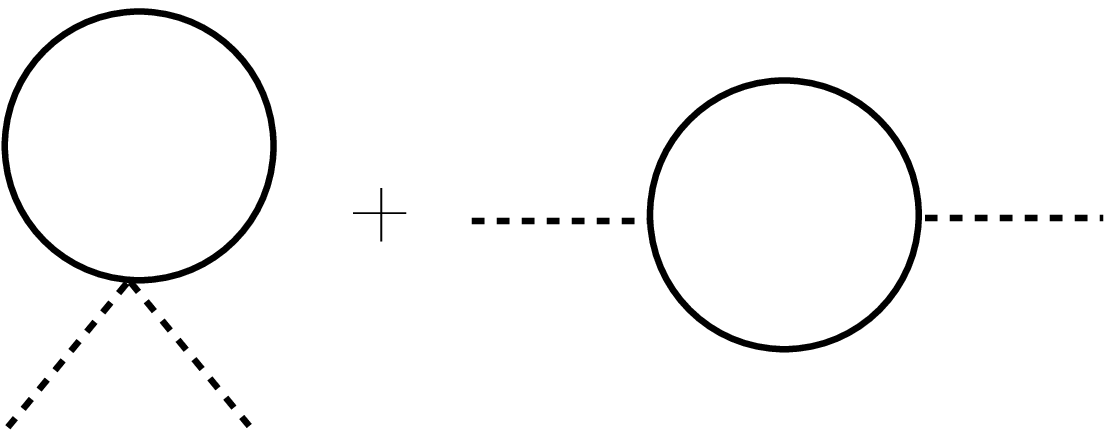}}\ \
&=& - \int \frac{dp}{2\pi}\,\frac{\delta X^2}{k^2+X^2} +
\frac{1}{2}\int \frac{dp}{2\pi}\,\frac{\delta X(k)\delta
X(-k)}{(p^2+X^2)((p+k)^2+X^2)}\,4X^2 \nn\\ &=& -\frac{k^2 \delta
X(k)\delta X(-k)}{8X^3}\ee
These corrections can be absorbed into a one-loop finite
renormalization of the gauge coupling constant. The low-energy
dynamics of the vector multiplet scalars is described by the
sigma-model
\be L_{\rm bosonic}=f(X)\dot{\vec{X}}{}^2\label{lbose}\ee
with
\be f(X)=\frac{1}{2g^2}+\frac{N}{4X^3}\label{f}\ee
A non-renormalization theorem due to Diaconescu and Entin
\cite{diac} shows that \eqn{lbose} is most general form of the
bosonic effective action with two derivatives, consistent with
${\cal N}=(4,4)$ supersymmetry and $Spin(5)$ invariance.

\para
The coulomb branch metric \eqn{f} exhibits a semi-infinite throat
as we approach the origin at $X=0$. Protected by the
non-renormalization theorem of \cite{diac}, we may extrapolate the
effective action to the regime $g^2N \gg X^3$ where we focus on
the throat region. Here we may drop the constant term in $f(X)$,
and the Coulomb branch metric becomes
\be f(X)=\frac{N}{4X^3}\label{newf}\ee
The modes propagating in this throat regime are decoupled in field
space from those in the asymptotic region of the Coulomb branch.
It was shown some years ago by Berkooz and Verlinde \cite{berkver}
that the correct interpretation of this throat region is a
description of the flavor singlet sector of the Higgs branch
quantum mechanics, through the map \eqn{map}. This now explains
our strange starting point: flavour singlet excitations of the
hypermultiplets have been recast as excitations of $\vec{X}$ in
the Coulomb branch throat. The effective bosonic action
\eqn{lbose} with $f(X)$ given by \eqn{newf} also describes the
dynamics of the D0-brane in other regimes of the string coupling,
as we now describe.

\subsection{Berry Connection as the Spin Connection}

We have seen that the strong coupling limit of the quantum
mechanics is described by a non-relativistic particle moving in
the curved background $ds^2 = f(X)d\vec{X}^2$. As in Section 2, we
are interested in states that transform in the $({\bf 4},{\bf 2})$
of the $Spin(5)\times SU(2)_R$ R-symmetry. These correspond to
particles carrying spin in the gravitational background. The spin
of a particle moving in a curved background undergoes parallel
transport. This holonomy is Berry's phase.

\para
The Lagrangian describing spinnning particles is the fermionic
completion of \eqn{lbose}. It was presented in \cite{diac},
\be L_{\rm fermi} = if(\bar{\Lambda}D_t{\Lambda}+
D_t{\bar{\Lambda}}\,\Lambda) -\ft12
R_{\alpha\beta\gamma\delta}\Lambda^\alpha\bar{\Lambda}^\beta
\Lambda^\gamma\bar{\Lambda}^\delta \label{lfermi}\ee
The final term is the usual sigma-model four-fermi coupling to the
Riemann tensor for the conformally flat metric $ds^2 = fdX^2$
(written here in the non-coordinate basis).

\para
The nature of the excited states is rather different at weak and
strong coupling. At weak coupling, they correspond to a
string-anti-string pair stretched between the D0-brane and
D4-branes. At strong coupling, the $D0-D4$ strings have been
integrated out in their ground state, and the first excited state
arises from a $D0-D0$ providing spin for the D0-brane. The
excitation spectrum comes from the Riemann tensor term in
\eqn{lfermi} which, after a little algebra, the Riemann tensor
becomes \cite{diac}
\be R_{\alpha\beta\gamma\delta}\Lambda^\alpha\bar{\Lambda}^\beta
\Lambda^\gamma\bar{\Lambda}^\delta = \left( f_{,\,\mu\nu}-\ft12
f^{-1}f_{,\,\mu}f_{,\,\nu}\right)
(\bar{\Lambda}\Gamma^\mu\Lambda\,\bar{\Lambda} \Gamma^\nu\Lambda +
\bar{\Lambda}\Gamma^\mu\bar{\Lambda} \,
\Lambda\Gamma^\nu\Lambda)\nn\ee
where $f_{,\,\mu}=\partial f/\partial X^\mu$. Quantization
requires us to impose the anti-commutator relations
$\{\bar{\Lambda},\Lambda\}\sim f^{-1}$, after which the four fermi
coupling gives rise to interactions which scale as $f^{-2}
f^{\prime\prime}$ and $f^{-3}f^{\prime\,2}$: these scale as $X/N$.

\para
Typically in an effective Lagrangian for quantum mechanics, one
may read off the induced Berry's connection from terms of the form
$\dot{X}^\mu A_\mu$ in the action \cite{japanese,moody}. Our
Lagrangian is no exception. Berry's phase appears in the fermionic
covariant derivative, which includes the spin connection governing
parallel transport
\be  D_t\Lambda^\alpha = \dot{\Lambda}^\alpha+ i
\dot{X}^\mu(\omega_\mu)^\alpha_{\ \beta}\Lambda^\beta\ee
A simple calculation of the spin connection for the conformally
flat metric $ds^2=f d\vec{X}{}^2$ yields
\be \omega_\mu = \frac{3}{2}\frac{{X}^\nu}{X^2}\Gamma_{\nu\mu} \ee
This connection is to be compared with the weak coupling result
\eqn{yy}. These represent the Berry connection of the quantum
mechanics for the lowest lying states in the $({\bf 4},{\bf 2})$
representation of $Spin(5)\times SU(2)_R$ at weak and strong
coupling respectively. The functional form is dictated by $SO(5)$
symmetry. Note that at weak coupling, the Berry connection was
reducible, as shown in the gauge \eqn{yy0} where it manifestly
mixes pairs of states. At strong coupling it mixes all four
states. At intermediate coupling, one expects further functional
dependence of the connection on the dimensionless ratio
$g^2N/X^3$. It may be that this function interpolates continuously
between the weak coupling coefficient of 1, and the strong
coupling coefficient of $3/2$. However, we cannot rule out the
possibility that level crossing occurs at some critical value such
that the states cannot be unambiguously transformed into each
other.

\subsection{The View from the D4-Brane}

We end this section by redressing the above calculation in
slightly different clothing. Instead of taking the decoupling
limit of the D0-brane, would could instead take the decoupling
limit of the D4-branes. The D4-brane coupling constant is $e^2 =
(2\pi)^2 g_s\sqrt{\alpha^\prime}$, and the decoupling limit for
the $d=4+1$ dimensional $U(N)$ gauge theory requires us to take
$g_s \rightarrow \infty$ and $\alpha^\prime \rightarrow 0$,
keeping $\vec{X}$ and $e^2$ fixed.

\para
When $e^2NX \ll 1$, the D4-brane theory is weakly coupled. In this
regime, the D0-brane is described by a Yang-Mills instanton inside
the D4-brane \cite{doug}. In $d=4+1$ dimensions, the instanton is
a solitonic particle. The low-energy dynamics of this soliton
coincides, via the ADHM construction, with the Higgs branch of the
D0-brane quantum mechanics. The $U(N)$ singlet sector of the
soliton dynamics is therefore well captured by the the Coulomb
branch throat \eqn{lbose}.

%(Note that, although not always well advertised, the agreement of
%the metric on the D0-brane Higgs branch and metric on the
%instanton moduli space is essentially due to a non-renormalization
%theorem which, in this case, is the constraint imposed by
%hyperKahlerity).

\para
The strong coupling limit of the $d=4+1$ dimensional gauge
theory is captured by the near-horizon limit of the D4-branes
\cite{imsy}. The type IIA supergravity solution describing $N$
coincident D4-branes is given by,
\be ds^2 = H(R)^{-1/2} d{\bf R}^{1,4} + H(R)^{+1/2}d{\bf
R}^5\ee
where the function $H(R)=1+g_sN {\pi\alpha^{\prime\,3/2}}/{R^3}$
is harmonic on the transverse ${\bf R}^5$. This metric is
accompanied by the dilaton $e^{-2(\phi-\phi_\infty)} = H^{1/2}$
and the RR 5-form $C_{0,\ldots,4}=-\ft12(H^{-1}-1)$. In the strong
coupling limit  $e^2NX \gg 1$, with $R=2\pi\alpha^\prime X$, we
may drop the "1" in the harmonic function $H$, and the physics of
the $d=4+1$ dimensional gauge theory is captured by the IIA
supergravity background. We also require $1\ll e^2NX \ll N^{4/3}$
to ensure that the dilaton does not blow up.

\para
The non-relativistic dynamics of the D0-brane moving in in the
background of the D4-brane is once again governed by the action
\eqn{lbose}. (The varying dilaton means that the D0-brane mass
$M_{D0}= e^{-\phi}/\sqrt{\alpha^\prime}$ is position dependent,
providing an extra factor of $H^{1/2}$ to augment the $H^{1/2}$
from the metric). Thus in the $d=4+1$ dimensional field theory,
the weak and strong coupling dynamics of the soliton coincide:
this, of course, is guaranteed by the non-renormalization theorem
\cite{diac}.

\para
This perspective, coupled with the topic of this paper, suggests a
novel way of extracting geometry from gauge theory. One looks at a
heavy, solitonic object in the field theory and, at weak coupling,
studies the semi-classical Berry connection for the spinning
states of the soliton as suitable operators, related to
coordinates in the gravitational dual, are varied. This Berry
connection is then interpreted as the spin connection of the dual
geometry. This is very much analogous to the view of  emergent
geometry through the eyes of an instanton \cite{dhkmv,hengyu}.
Indeed, for the D0-D4 system, this perspective does not offer
anything new over and above the computations of
\cite{diac,berkver}. However, placing this new spin on the soliton
spin may provide a useful emphasis in studying other systems.

\section*{Appendix A: Useful Matrices}

\setcounter{section}{1}\setcounter{equation}{0}
\renewcommand{\theequation}{\Alph{section}.\arabic{equation}}

The discussion in the main text has, for the most part, been
convention independent. One exception is the explicit connection
for the non-Abelian Berry's phase \eqn{yang} which assumes both a
specific basis for the Euclidean 5d gamma matrices $\vec{\Gamma}$,
as well as a basis of states. Our choice of the former is,
\be &\Gamma_1 = \scriptsize{\left(\begin{array}{cccc}
0 & 0 & 0 & -1\\
0 & 0 & 1 & 0\\
0 & 1 & 0 & 0\\
-1 & 0 & 0 & 0  \end{array} \right)} \,,\qquad \Gamma_2 =
{\left(\begin{array}{cccc}
0 & 1 & 0 & 0\\
1 & 0 & 0 & 0\\
0 & 0 & 0 & 1\\
0 & 0 & 1 & 0  \end{array} \right)}\,,\qquad \Gamma_3  =
{\left(\begin{array}{cccc}
0 & -i & 0 & 0\\
i & 0 & 0 & 0\\
0 & 0 & 0 & i\\
0 & 0 & -i & 0  \end{array} \right)}&\nn\\  & \Gamma_{4} =
\scriptsize{\left(\begin{array}{cccc}
0 & 0 & 0 & -i\\
0 & 0 & i & 0\\
0 & -i & 0 & 0\\
i & 0 & 0 & 0  \end{array} \right)} \,,\qquad \Gamma_{5}  =
{\left(\begin{array}{cccc}
1 & 0 & 0 & 0\\
0 & -1 & 0 & 0\\
0 & 0 & 1 & 0\\
0 & 0 & 0 & -1  \end{array} \right)}& \ee
The Berry connection \eqn{yang} also employs the self-dual 't
Hooft symbols $\eta$. These are three $4\times 4$ matrices
satisfying the $SU(2)$ commutation relations, given by
%
%\be \eta^m{}_{\mu\nu} &= \epsilon_{m\mu\nu4} +
%\delta_{m\mu}\delta_{4\nu} - \delta_{m\nu}\delta_{4\mu}\ee
%
%Explicitly, these are given by,
%
\be {\eta}^1=\scriptsize{\left(\begin{array}{cccc}0 &0 &-1 & 0
\\0 & 0& 0 & 1
\\ 1& 0 & 0&0 \\ 0 &-1 &0 &0 \end{array}\right)} \ \ \ ,\ \ \ \
{\eta}^2={\left(\begin{array}{cccc}0 &-1 &0 & 0 \\1 & 0& 0 & 0
\\ 0& 0 & 0&-1 \\ 0 & 0 &1 &0 \end{array}\right)}\ \ \ \ ,\ \ \ \
{\eta}^3={\left(\begin{array}{cccc}0 &0 &0 & 1 \\0 & 0& 1 & 0
\\ 0& -1 & 0&0 \\ -1 &0 &0 &0 \end{array}\right)}
\label{thooft}\ee
The Yin monopole is written in terms of the anti-self-dual 't
Hooft symbols $\bar{\eta}$
%
% \be \bar\eta^m{}_{\mu\nu} &= \epsilon_{m\mu\nu4}
%- \delta_{m\mu}\delta_{4\nu} + \delta_{m\nu}\delta_{4\mu} \ee
%
%which read
%
\be \bar{\eta}^1=\scriptsize{\left(\begin{array}{cccc}0 &0 &-1 & 0
\\0 & 0& 0 & -1
\\ 1& 0 & 0&0 \\ 0 &1 &0 &0 \end{array}\right)} \ \ \ ,\ \ \ \
\bar{\eta}^2={\left(\begin{array}{cccc}0 &1 &0 & 0 \\-1 & 0& 0 & 0
\\ 0& 0 & 0&-1 \\ 0 &0 &1 &0 \end{array}\right)}\ \ \ \ ,\ \ \ \
\bar{\eta}^3={\left(\begin{array}{cccc}0 &0 &0 & 1 \\0 & 0& -1 & 0
\\ 0& 1 & 0&0 \\ -1 &0 &0 &0 \end{array}\right)}
\label{antithooft}\ee

\section*{Appendix B: Berry's Phase and the Yang Monopole}
\setcounter{section}{2}\setcounter{equation}{0}

In this appendix we give the full details of the computation of
the $SU(2)$ Berry connection \eqn{yang}. Since the bosons $\phi$
and $\tilde{\phi}$ play no role in Berry's phase we disregard them
in the following calculation and consider only the fermions. We
start with the four-dimensional Hilbert space ${\cal H}_4$ spanned
by the states $\bar{\Psi}_\alpha |\,0\rangle$. The fermionic
Hamiltonian on this space acts as
\be H_F=\vec{X}\cdot\vec{\Gamma}\ee
As in the main text, we introduce the projection operators onto
the eigenspaces of $H_F$,
\be P_{\pm} = \frac{1}{2}\left(\bbbone \pm
\frac{\vec{X}}{X}\cdot\vec{\Gamma} \right) \ee
To compute the Berry connection, we must first choose a gauge.
This means picking an arbitrary choice of basis vectors for the
ground states in $P_-{\cal H}_4$ for each value of $\vec{X}$. We
choose the basis of (un-normalized) states,
\be |\,1\rangle = P_-\bar{\Psi}_2|\,0\rangle \ \ {\rm and}\ \ \
|\,2\rangle = P_- \bar{\Psi}_4|\,0\rangle\ee
which are valid everywhere except along the $\vec{X}=(0,0,0,0,-X)$
axis where the ground states are perpendicular to both
$\bar{\Psi}_2|\,0\rangle$ and $\bar{\Psi}_4|\,0\rangle$. As a
result, the Berry connection we derive will have a Dirac string
singularity  along this axis. To be more explicit, we introduce
standard spherical polar coordinates on ${\bf R}^5$,
\be
X^1 &=& X \sin\theta_1 \sin\theta_2 \sin\phi_2\nonumber\\
X^2 &=& X \sin\theta_1 \cos\theta_2 \cos\phi_1\nonumber\\
X^3 &=& X \sin\theta_1 \cos\theta_2 \sin\phi_1\\
X^4 &=& X \sin\theta_1 \sin\theta_2 \cos\phi_2\nonumber\\
X^5 &=& X \cos\theta_1\nonumber \ee
Then, in the basis of states $\bar{\Psi}_\alpha|\,0\rangle$, the
two normalized ground states are given by
\be
 |\,1\rangle = \frac{1}{\sqrt{2(1+\cos\theta_1)}}
 \scriptsize{\left(\begin{array}{c}
  -\cos\theta_2\sin\theta_1 e^{-i\phi_1} \\ 1+\cos\theta_1\\ i
  \sin\theta_1\sin\theta_2 e^{i\phi_2} \\ 0 \end{array}\right)} \ , \
  \ \
 |\,2\rangle = \frac{1}{\sqrt{2(1 + \cos\theta_1)}}
\scriptsize{\left(\begin{array}{c}
 i\sin\theta_1\sin\theta_2 e^{-i\phi_2} \\ 0 \\
 -\cos\theta_2\sin\theta_1 e^{i\phi_1} \\ 1 + \cos\theta_1\end{array}\right)}
\nn\ee
With this explicit parameterization of the states in hand, we may
now compute the non-Abelian Berry connection defined by
\be (A_{\mu})_{ab} = i\langle a |\frac{\partial}{\partial
X^\mu}|b\rangle \ee
A straightforward computation yields the one-forms $A_{ab}\equiv
(A_\mu)_{ab}dX^\mu$,
\be A_{11}&=&-A_{22} =  \sin^2\ft{\theta_1}{2}\left(
\cos^2\theta_2\, d\phi_1 - \sin^2 \theta_2\, d\phi_2  \right)\nn\\
%&=&\frac{1}{2X(X-X^5)}( X^1\, d X^2 - X^2\, dX^1 + X^3\, dX^4
%- X^4\, dX^3)\\
A_{12}&=& A_{21}^\star =
 \sin^2\ft{\theta_1}{2}\,e^{i(\phi_1 - \phi_2)} \left( d \theta_2
 -
 i\cos\theta_2 \sin\theta_2(d\phi_1 + d\phi_2)  \right)
% &=&\frac{-i}{2x(x-x_5)}\Biggl( x_4\dd x_1 - x_1 \dd x_4
%+ x_3 \dd x_2 - x_2 \dd x_3 \nonumber \\
% &=& + i\left( x_3 \dd x_1 - x_1 \dd x_3 + x_2 \dd x_4 - x_4 \dd x_2
% \right)\Biggr)
\ee
The relationship $A_{11}=-A_{22}$ reflects the fact that the
central $U(1)$ part of the $U(2)$ connection is trivial. The
states undergo only a $SU(2)$ holonomy. Translating back to
Euclidean coordinates, we can express the Berry connection in the
form given in the main text,
\be (A_\mu)_{ab} =
\frac{-X^\nu}{2X(X+X_5)}\,\eta^m_{\mu\nu}\,\sigma^{m}_{ab} \qquad
\mu = 1,2,3,4 \ \ \ \ ,\ \ \ \ A_5 =0 \ee
where $\eta_{\mu\nu}^m$ are the 't Hooft matrices defined in
Appendix A and $\sigma ^m$ are the Pauli matrices.

\subsubsection*{Yin Monopole}

The calculation for the sector $\bar{\Psi}_\alpha\bar{\Psi}_\beta
\bar{\Psi}_\gamma|\,0\rangle$ proceeds in a similar manner. The
operator $P_+$ now projects onto the relevant states. Working in
the basis ${}^*\bar{\Psi}_\alpha|\,0\rangle =J_\alpha{}^\beta
\epsilon_{\beta\gamma\delta\rho}\bar{\Psi}_\gamma\bar{\Psi}_\delta
\bar{\Psi}_\rho |\,0\rangle$, we take the normalized ground states
to be
\be |\,\tilde{1}\rangle = P_+{}^\star\bar{\Psi}_1|\,0\rangle \ \
{\rm and}\ \
%=\frac{1}{\sqrt{2(1+\cos\theta_1)}}
%\scriptsize{\left(\begin{array}{c} 1 + \cos\theta_1 \\
%\cos\theta_2\sin\theta_1\,e^{i\phi_1} \\ 0 \\ i
%  \sin\theta_1\sin\theta_2\,e^{-i\phi_2}\end{array}\right)} \nn\\
|\tilde{2}\rangle = P_+{}^\star \bar{\Psi}_3|\,0\rangle
%=\frac{1}{\sqrt{2(1+\cos\theta_1)}}
%\scriptsize{\left(\begin{array}{c}
% 0 \\ i
% \sin\theta_1\sin\theta_2\,e^{i\phi_2} \\ 1 + \cos\theta_1 \\ \cos\theta_2
% \sin\theta_1\,e^{-i\phi_1}\end{array}\right)}
\ee
which again go bad along the negative $X^5$ axis. We may now
compute the Berry connection in this sector: $\tilde{A}_{ab} =
i\langle \tilde{a} |\frac{\partial}{\partial
X^\mu}|\tilde{b}\rangle\,dX^\mu$. It is given by
\be \tilde{A}_{11} &=& -\tilde{A}_{22} =
\sin^2\ft{\theta_1}{2}\,\left(
-\cos^2\theta_2 d \phi_1 - \sin^2 \theta_2 d \phi_2  \right)\nn \\
\tilde{A}_{12} &=& \tilde{A}_{21}^\star =
 \sin^2\ft{\theta_1}{2}\,e^{-i(\phi_1 + \phi_2)} \left(-d \theta_2 - i
 \cos\theta_2\sin\theta_2(d\phi_1 - d\phi_2)  \right)
\ee
Returning once more to Cartesian coordinates, we have
\be (\tilde{A}_\mu)_{ab} = (A_\mu)^\star_{ab} =
\frac{-X^\nu}{2X(X+X_5)}\,\bar{\eta}^m_{\mu\nu}\,\sigma^m_{ab} \ee
where $\bar\eta^m$ are the anti-self-dual 't Hooft symbols defined
in \eqn{antithooft}. The Berry connection in this sector is
therefore the anti-Yang, or Yin, monopole.

\subsection*{Acknowledgements}

We would like to thank Ofer Aharony, Nick Dorey, Juan Maldacena,
Nick Manton and Paul Townsend for useful comments and discussions.
D.T. thanks the Isaac Newton Institute for hospitality during the
completion of this work. C.P. is supported by an EPSRC
studentship. J.S. is supported by the Gates Foundation and STFC.
D.T. is supported by the Royal Society.

 \end{document}